# Structural, optical, and thermal properties of BN thin films grown on diamond via pulsed laser deposition

Abhijit Biswas,[1,*] Gustavo A. Alvarez,[2] Tao Li,[3] Joyce Christiansen-Salameh,[2] Eugene Jeong,[2] Anand B. Puthirath,[1] Sathvik Ajay Iyengar,[1] Chenxi Li,[1] Tia Gray,[1] Xiang Zhang,[1] Tymofii S. Pieshkov,[1,4] Harikishan Kannan,[1] Jacob Elkins,[1] Robert Vajtai,[1] A. Glen Birdwell,[5] Mahesh R. Neupane,[5] Elias J. Garratt,[5] Bradford B. Pate,[6] Tony G. Ivanov,[5] Yuji Zhao,[3] Zhiting Tian,[2,*] and Pulickel M. Ajayan[1,*]

**AFFILIATIONS**

[1]Department of Materials Science and Nanoengineering, Rice University, Houston, TX, 77005, USA

[2]Sibley School of Mechanical and Aerospace Engineering, Cornell University, Ithaca, NY 14853, USA

[3]Department of Electrical and Computer Engineering, Rice University, Houston, TX, 77005, USA

[4]Applied Physics Graduate Program, Smalley-Curl Institute, Rice University, Houston, TX, 77005, USA

[5]DEVCOM Army Research Laboratory, RF Devices and Circuits, Adelphi, MD 20783, USA

[6]Naval Research Laboratory, Chemistry Division, Washington, DC, 20375 USA

[*]Corresponding Authors: **01abhijit@gmail.com, zhiting@cornell.edu, ajayan@rice.edu**

**Abstract**

Heterostructures based on ultrawide-bandgap (UWBG) semiconductors (bandgap >4.0 eV), such as boron nitride (BN) and diamond, hold significant importance for the development of high-power electronics in the next generation. However, achieving in-situ hetero-epitaxy of BN/diamond or vice versa remains exceptionally challenging due to the complex growth kinetics involved. In this study, we have grown BN thin film on (100) single crystal diamonds using pulsed laser deposition and investigated its structural, magnetic, optical, and thermal properties. The structural analyses confirmed the growth of BN films, which exhibited diamagnetic behavior at room temperature. Notably, the film demonstrated anisotropic refractive index characteristics within the visible-to-near-infrared wavelength range. The room temperature cross-plane thermal conductivity of BN is 1.53 (± 0.77) W/mK, while the thermal conductance of the BN/diamond interface is 20 (± 2) MW/m$^2$K. These findings have significant implications for a range of device applications based on UWBG BN/diamond heterostructures.

# I. INTRODUCTION

Epitaxial integration of ultrawide-bandgap (UWBG) semiconductors (bandgap > 4.0 eV) represent a revolutionary new arena in materials research since they show excellent promises for high-performance electronics.[1-4] Typical existing UWBG semiconductors are diamond, boron nitride (BN), AlGaN/AlN, and $\beta$-$Ga_2O_3$.[1-4] Since the figure-of-merit (FOM) of a device scales with the bandgap,[1] these UWBG semiconductors show compelling advantages over their WBG counterpart (e.g. GaN and SiC), especially for high-power radio frequency (RF) electronics, deep-ultraviolet (UV) optoelectronics, quantum information, thermal management, and for harsh-environmental applications.[5, 6] Despite tremendous potentials, however, the research in UWBG semiconductors is still premature, as integrating UWBG semiconductors with smooth and trapped-charges-free interfaces remains very challenging.[1] It's extremely important for the improved device performances with the reduced interfacial scattering, to supersede the WBG technology.[1]

According to the Baliga-figure-of-merit (BFOM), diamond and cubic boron nitride (c-BN), the two strongest materials, show great potentials in devices, with high breakdown voltage (~$10^4$ V) and lower loss per switching cycle.[1] Diamond (bandgap $E_g$ ~5.5 eV) and c-BN ($E_g$ ~6.2 eV) are structurally, chemically, thermally, and electrically compatible [**Fig. 1(a)**].[7] Electronically, both *p*- and *n*-type doping are possible in c-BN, while only *p*-type is achievable for diamond.[1-3] A high-mobility field-effect transistor (FET) is thus feasible by making heterojunctions of *n*-type c-BN and *p*-type diamond.[1-3] Therefore, extensive efforts are ongoing for the *in-situ* fabrication of electronic quality UWBG c-BN/diamond thin film heterostructures or vice-versa.[8-17]

In-situ epitaxial thin film growth of single crystalline electronic quality pure or doped c-BN on structurally lattice-matched diamond and vice-versa is the holy grail for UWBG devices. For BN polymorphs, the most stable phase is the hexagonal two-dimensional BN structure (2D h-BN, $E_g$ ~5.9 eV), whereas 3D c-BN is the metastable phase and forms only at high-temperature and pressure [**Fig. 1(a)**].[7, 18] Depending on the pressure and temperature during the growth, BN forms various polymorphs and thus the mixed phase films.[8-17] Zhang *et al.*, wrote comprehensive reviews on the growth related issues of mixed phase BN film on Si and Diamond.[8,12] Shtansky *et al.*, made efforts to provide the mechanism of mixed phase BN growth on Si.[10] Narayan *et al.*, showed the direct conversion of h-BN into nano-ordered c-BN by nanosecond pulsed laser melting.[14] Hirama *et al.*, grew films with abrupt interface on diamond and observed that high temperature is important for c-BN growth.[15,16] Therefore, as an alternative, ex-situ mechanically transferred h-BN flakes on single crystal diamond (SCD) has been used as a gate dielectric in some diamond-based devices.[19, 20] These diamond FETs show moderate mobility, due to defects at the h-BN/SCD interface. Hence, there is a desperate need for the *in-situ* growth of defect-free thin layer of BN on (100) SCD with a clean interface. Of greater significance, the measurement of relevant optical and thermal properties holds importance for the advancement of UWBG diamond-based high power electronics.

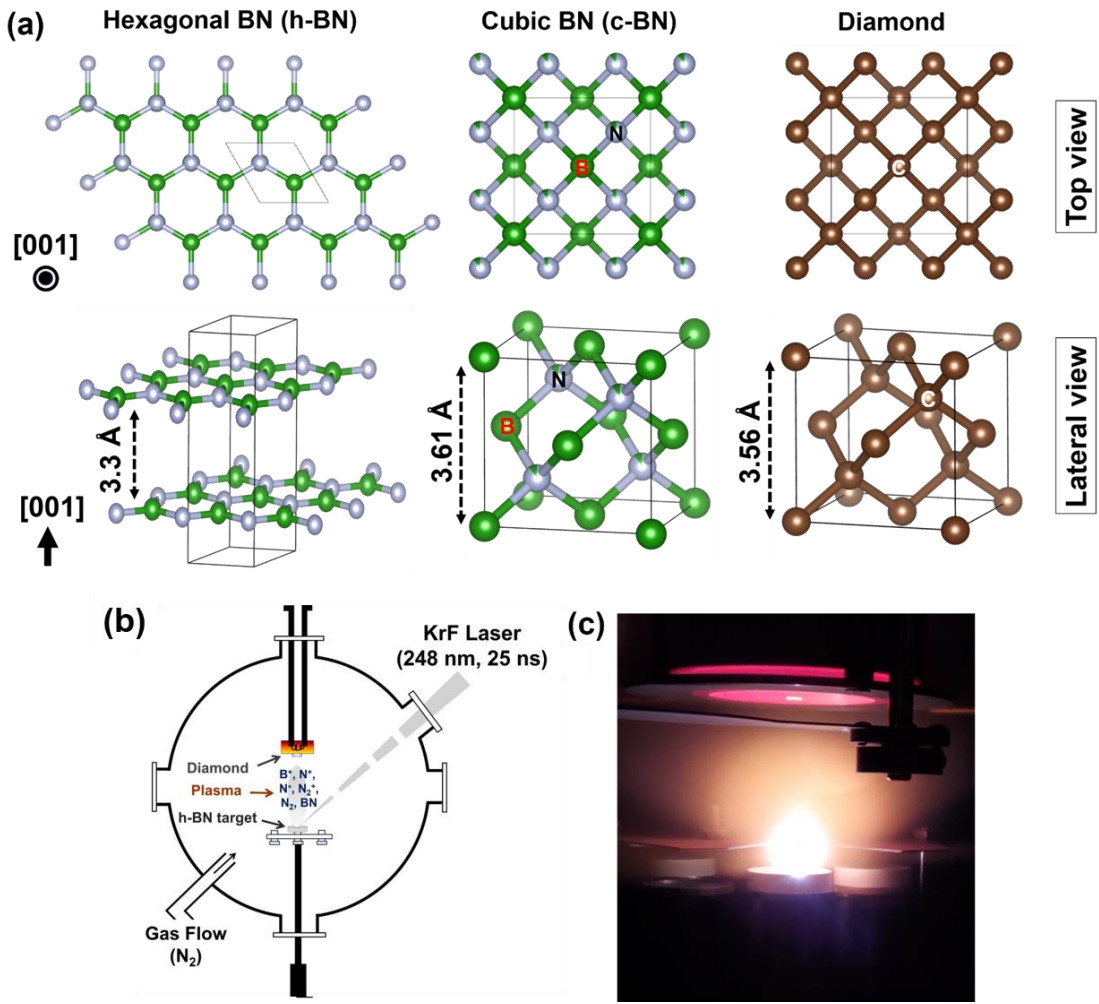

**FIG. 1.** (a) Structural compatibility of hexagonal, cubic boron nitride and diamond (upper panel: top view; lower panel: lateral view). (b) Schematic of pulsed laser deposition process. (c) Image captured during the real-time growth showing laser ablated plasma travelling towards the diamond substrate (kept on a hot substrate holder facing downward). Structures are drawn using VESTA software.

In literature, several efforts have been made to grow BN, specially c-BN films, by chemical vapor deposition (CVD) or molecular beam epitaxy (MBE).[8-17] However, it forms mixed phase nanocrystalline BN, with interfacial defects (e.g. stacking faults), because of the complex growth kinetics. It has been predicted that additional energy is required to adatoms during the growth.[8, 9] Thus the use of a highly energetic non-equilibrium pulsed laser deposition (PLD) might be an appropriate process. The advantages of PLD is the use of a dense target (single or polycrystalline) which is ablated by a highly energetic (2-5 eV) UV-pulsed laser, in the presence of partial gas pressure. PLD also offers the faithful transfers of elements from the target-to-substrate surface in the form of plasma (in a stoichiometric ratio), endowed with the transfer of high-energy radicals and ions ($B^+$, $N^*$, $N_2^*$, $B^+$, $N^+$, BN), with a kinetic energy ~10-100 eV [**Figs. 1(b)** and **1(c)**].[21] These ablated energetic species travel to the substrate within a few µs, and coalesce to form a structure similar to the substrate and thereby result in epitaxial film growth.

Here, we have grown BN thin film on (100) SCD substrates by PLD and performed structural, magnetic, optical, and thermal characterizations. Structural characterizations confirm the growth of BN. The films show diamagnetic behavior, have an anisotropic refractive index within the visible-to-near-infrared wavelength range, and excellent cross-plane thermal conductivity of 1.53 ($\pm$ 0.77) W/mK, and interfacial thermal conductance of 20 ($\pm$ 2) MW/m²K at room temperature. These *in-situ* grown BN films on diamond by PLD with the exhibition of various functionalities might be useful for the optoelectronics and critical thermal management applications.

## II. RESULTS AND DISCUSSION

### A. Diamond substrate characterizations

For depositions, we used commercially available normal-grade (100) SCD substrates (SKU 145-500-0549, size: 3× 3 mm²), purchased from Element Six (UK). Since the surface of diamond is crucial,[1, 7] we first characterized the quality of SCD. We performed X-ray diffraction (XRD) by using the Rigaku SmartLab X-ray diffractometer, equipped with a monochromatic Cu Kα radiation source. Thin film XRD show the presence of only (400) peak throughout the two-theta ($\theta$-$2\theta$) scan range, with a full width at half maxima (FWHM) of ~0.03° [**Fig. 2(a)**]. We obtained the miscut angle of SCD by measuring the FWHM of (400) rocking curves at four-different azimuthal angles, which is found to be of ~1.21° [**Fig. 2(b)**]. The surface topography by atomic force microscopy (AFM) shows flat surface with a surface roughness of ~1.5 nm [inset of **Fig. 2(a)**]. X-ray photoelectron spectroscopy (XPS) (performed by using the PHI Quantera SXM scanning X-ray microprobe with a 1486.6 eV monochromatic Al Kα X-ray source and at 26 eV pass energy) of the as-received SCD show carbon peak along with the presence of oxygen and Si impurity [**Fig. 2(c)**]. Therefore, we annealed SCD substrate at ~750 °C and in nitrogen atmosphere ($P_{N2}$ ~100 mTorr) for an hour (inside the PLD chamber), which resulted in a notable reduction in both the oxygen and Si concentrations. Raman spectra (using the Renishaw inVia confocal microscope with the 532-nm laser as the excitation source), shows strong D-band (~1333 cm$^{-1}$), along with the nitrogen vacancy related peaks (NV$^0$ and NV$^-$), G-band and second-order Raman spectrum [**Fig. 2(d)**].[22] These vacancy related peaks are reduced after the high-temperature in-situ annealing. Therefore, it is a necessity to perform in-situ annealing of SCD substrates before thin film depositions.

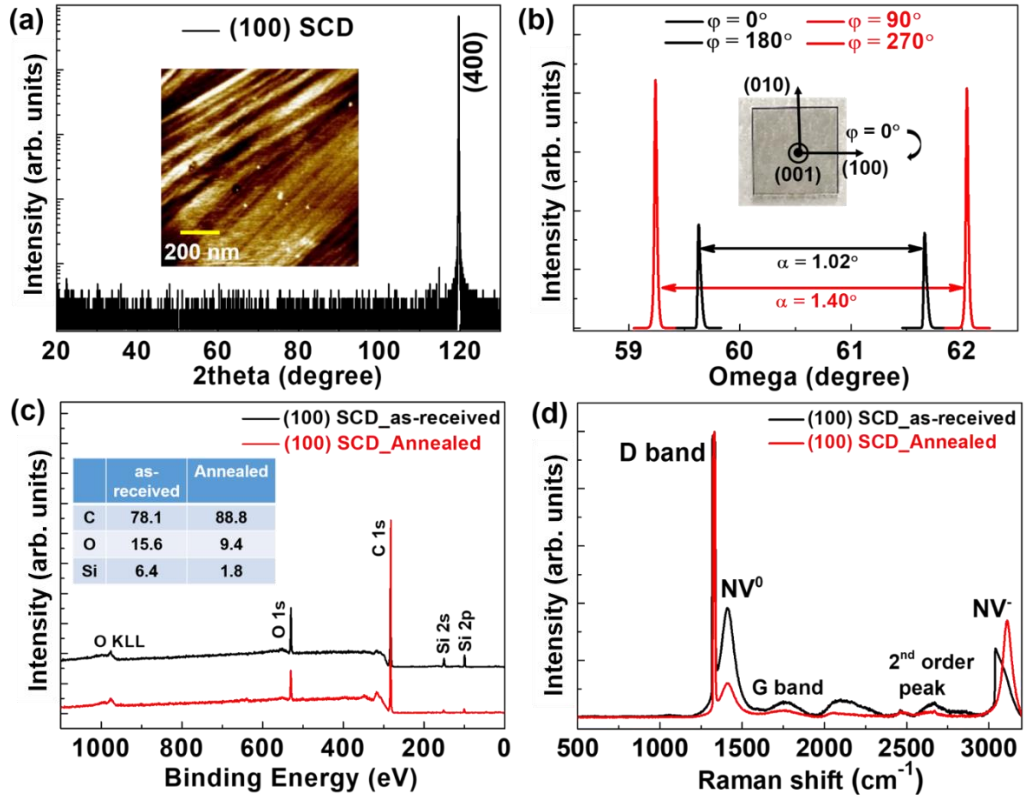

**FIG. 2.** (a) X-ray diffraction shows only the (400) peak. Inset shows the atomic force microscopy image with surface roughness of ~1.5 nm. (b) Rocking curve at four different azimuthal angles showing the average miscut angle of ~1.2°. Inset shows a diamond substrate (size ~3×3 mm$^2$). (c) X-ray photoelectron spectroscopy of as received and in-situ annealed diamond surface, showing that in-situ annealed substrate has a lower concentration of Si and oxygen. (d) Raman spectra show the characteristic D-band peak (~1330 cm$^{-1}$), corresponding to the vibration of the sp$^3$ diamond, along with the higher-orders and nitrogen vacancy related modes.

## B. Thin film growth and structural characterizations

BN thin films of thicknesses ~20 nm are grown by using PLD (load-lock assisted high vacuum (~5×10$^{-9}$ Torr) chamber operating with a KrF excimer laser (248 nm wavelength and pulse width 25 ns). The films are grown by using the following conditions: growth temperature ~750 °C, nitrogen partial pressure (P$_{N2}$) ~100 mTorr, and laser fluency ~2.2 J/cm$^2$, target-to-substrate distance ~ 50 mm, and repetition rate of 5 Hz. For the ablation, we used a commercially available one-inch diameter h-BN target (American Element, 99.9%). Before and after each deposition, SCD and BN/SCD is pre and post-annealed at the same growth temperature and pressure for one hour. For the depositions, 2000 laser pulses are supplied. After growth, films are cooled down at ~20 °C/min.

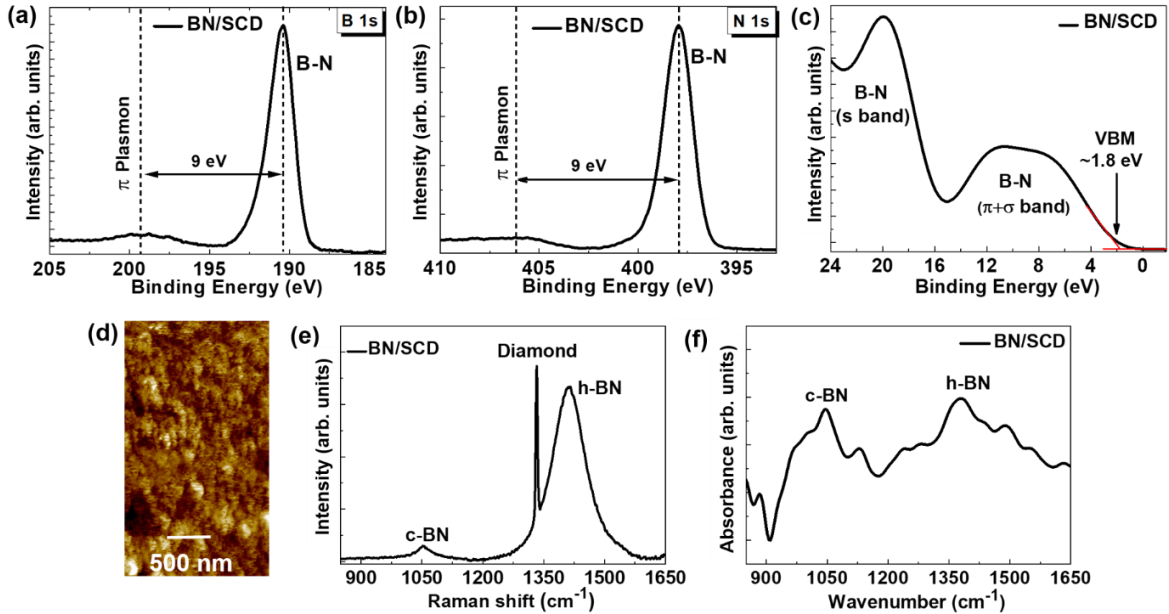

**FIG. 3.** (a), (b) X-ray photoelectron spectroscopy shows the characteristic B-N bonding with Plasmon peaks. (c) The XPS valence band spectra (VBS), with characteristics s-band and π+σ band. (d) Atomic force microscopy of BN surface. (e), (f) Raman and FTIR spectra of film.

We performed core-level XPS, which shows the presence of characteristic B-N bonding peaks at ~190.4 eV (B 1s-core) and ~397.9 eV (N 1s-core) [**Figs. 3(a)** and **3(b)**].[23, 24] In XPS, we also observed the Plasmon's peaks (at ~9 eV and ~25 eV apart from the main B-N peak), which is characteristic of h-BN.[25] The XPS-valence band spectra (XPS-VBS) was acquired by using the pass energy of 69 eV. The XPS-VBS show that the valence band maxima (VBM) is ~1.8 eV below the Fermi level ($E_F$) [**Fig. 3(c)**]. The films show a smooth surface with a roughness of ~2.88 nm [**Fig. 3(d)**]. In Raman spectra, we observed a peak at ~1053 cm$^{-1}$, that corresponds to c-BN transverse optical (TO) phonon mode [**Fig. 3(e)**].[8] Moreover, intense and broad peak within ~1350-1450 cm$^{-1}$ indicates the presence of $E_{2g}$ phonon mode of h-BN (along with NV$^0$ peak of SCD). Furthermore, Fourier-transform infrared (FTIR) spectra obtained by using the Nicolet 380 FTIR spectrometer by using a single crystal germanium window shows the TO phonon modes (both in plane and out-of-plane modes of c-BN and h-BN) [**Fig. 3(f)**].[8, 24]

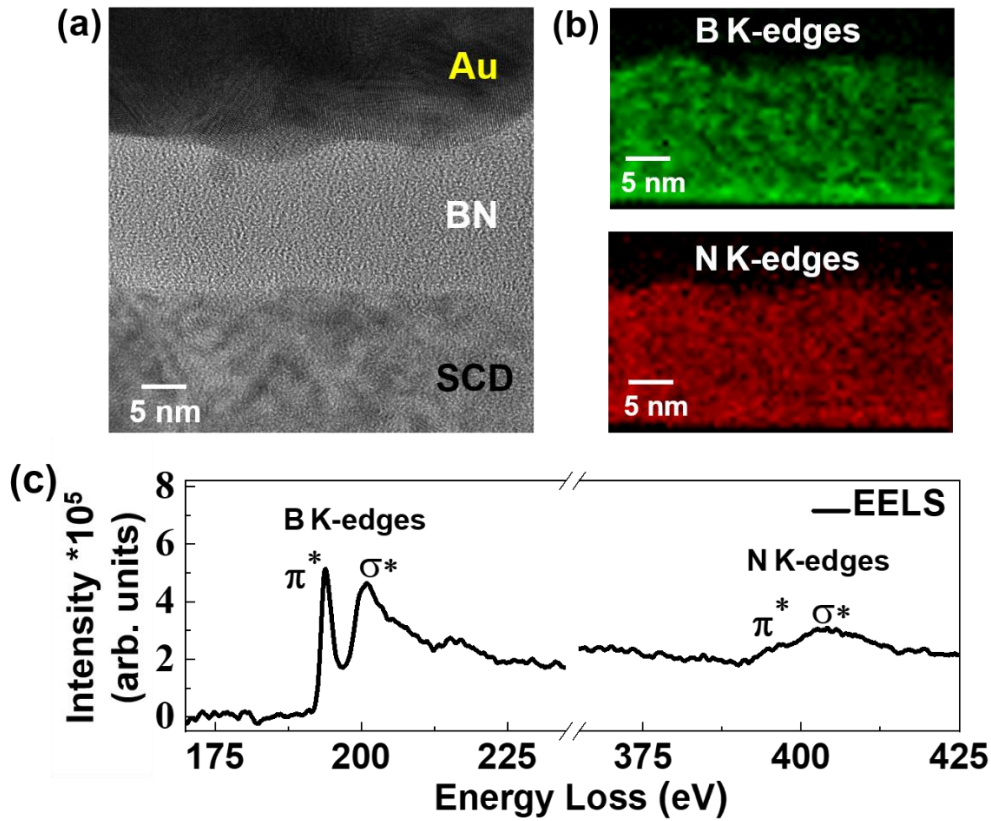

**FIG. 4.** (a) Cross-sectional high-resolution transmission electron microscopy image of BN film on SCD. (b) EELS elemental mapping shows the presence of both B and N. (c) EELS showing the bands corresponding to boron and nitrogen K-edges.

To get more insights about the crystallinity and phase of BN, cross-sectional high-resolution transmission electron microscopy (HRTEM) was conducted [**Fig. 4(a)**]. The obtained cross-sectional image clearly shows the presence of BN, which is further supported by electron-energy-loss spectroscopy (EELS) elemental mapping [**Fig. 4(b)**]. Corresponding B and N K-edges from the EEL spectra are also presented. [**Fig. 4(c)**]. The EELS spectra closely resembles the literature reports,[26, 27] indicating the presence of different phases of BN. The B-spectrum displays a distinct and narrow π* peak, reminiscent of h-BN. Additionally, a broader and slightly weaker σ* peak is also observed. The presence of both peaks suggests the coexistence of $sp^2$-bonded h-BN and a $sp^3$-bonded BN, which is probably the nucleation sites for c-BN.[27] The N-spectrum also resemble the feature of amorphous BN (*a*-BN) or c-BN, with a marginally intense π* peak than the broader and intense σ* peak.[26, 27] Furthermore, comparing the $sp^2$:$sp^3$ ratio in B-spectrum, it is evident that the volume fraction of $sp^2$ is slightly higher than the $sp^3$ BN (seen in the FTIR as well [**Fig. 3(f)**]).[26]

Considering the structural and chemical compatibility, c-BN is more favorable to grow on SCD, and thus extensive efforts are ongoing to understand the complex mixed phase growth of BN on SCD.[8-10] At ambient conditions, h-BN is the most thermodynamically stable phase. However, several theoretical predictions suggest that c-BN is stable at ambient conditions,[28, 29] making the growth kinetics complicated for pure c-BN phase film optimization. The huge kinetic barrier hinders the transformation of $sp^2$ h-BN (ionic) to $sp^3$-bonded (covalent) c-BN.[30] Therefore, it is thought that c-BN growth could be achieved by providing additional energy during the growth (e.g., high temperature growth kinetics, using ion-beam assisted growth, and post-growth thermal treatment).[8-10, 31, 32] However, these approaches develop additional stress, producing nanocrystalline films with defective interfaces.[33-35] Recently, magnesium (Mg) is used as a catalyst to reduce the energy barrier and promote the c-BN growth, however films remain as nanocrystalline phase.[36] We carried out growth at high temperatures and followed it with thermal annealing. Our method also resulted in the formation of BN, although it is likely to have a combination of different phases. Existing literatures have shown that non-crystalline, amorphous semiconductors (absence of long-range crystalline nature) possess exceptional functional properties, making them highly valuable in the field of electronics..[37, 38]

## C. Magnetic, optical, and thermal characterization of films

The magnetic properties of BN are also interesting.[39] Neither B nor N have any unpaired electrons, thus the single-crystal defect-free BN does not give any magnetic response. However, the incorporation of dopants (e.g., carbon or fluorine) or vacancies, results in a ferromagnetic response (formation of local moments) above room temperature.[39] Thus, it is worthwhile to explore the magnetic properties of BN films. We investigated the room temperature magnetic hysteresis loop of the film. Room temperature magnetic hysteresis (moment vs. magnetic field) measurement was performed using a SQUID-VSM-7 Tesla (Quantum Design, USA), within the magnetic field sweep range of ±5 Tesla. The film remains diamagnetic throughout the magnetic field scanning range [**Fig. 5(a)**], indicating that the grown BN film is free of defects or vacancies (also supports the XPS elemental scans showing the B:N ratio of ~1:1, precisely 35.4:35.2), within the detection limit of the SQUID instrument.

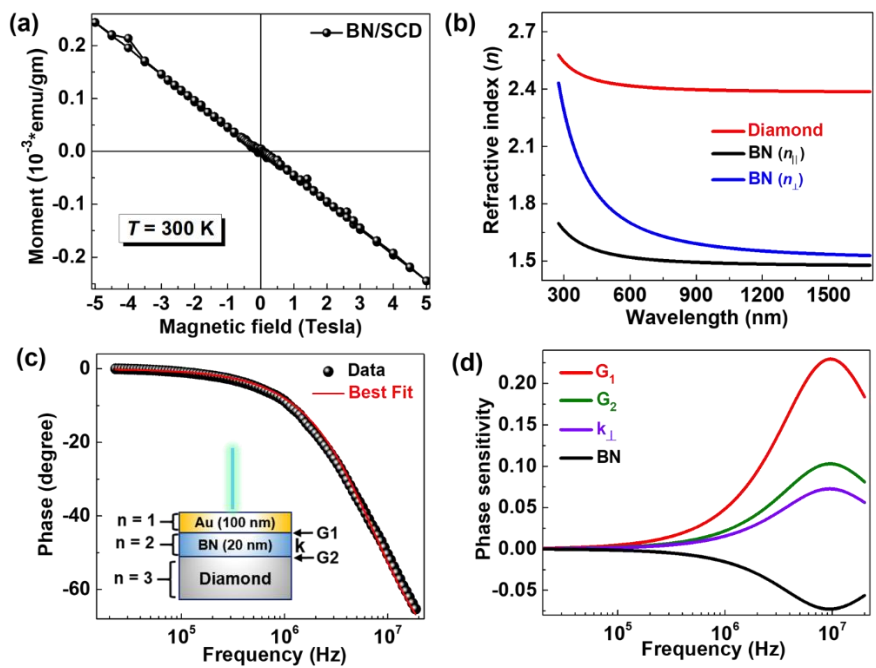

**FIG. 5.** (a) Room temperature magnetic hysteresis loop shows the diamagnetic behavior. (b) In-plane and out-of-plane refractive index of BN film and diamond within the visible-to-near-infrared wavelength. (c) Phase vs. frequency data for BN. Inset shows the multilayer sample model where each layer includes the volumetric heat capacity ($c_p$) cross plane ($k_\perp$) and in-plane ($k_\parallel$) thermal conductivity, layer thickness, and the thermal boundary conductance ($G_1$ and $G_2$). (d) Sensitivity analysis of the $G_1$ between gold and BN, $k_\perp$ of BN, $G_2$ between the BN and diamond.

We measured the optical refractive index (RI) of the film (wavelength range of 300-1600 nm). Variable angle spectroscopic ellipsometer (VASE) (M-2000 Ellipsometer, by J. A. Woollam Company) was used to measure the RI of pristine SCD (as a reference), followed by the RI of BN. The process involves illuminating the samples with light at four equally spaced incident angles ($\Phi$) ranging from 55° to 70° and collecting the reflected light to get the reflection coefficients for both s-polarized and p-polarized lights of the samples, and fitted to obtain the RI. For fitting, we constructed a two-layer model to represent the BN/SCD heterostructure, with an anisotropic uniaxial Cauchy model applied to the top layer.[38] This approach resulted in a mean square error (MSE) of ~13.6 and a physical normal dispersion across the entire wavelength regions. For BN film, we obtained the RI of ~1.5-1.7 (in plane, $n_\parallel$) and ~1.6-2.4 (out-of-plane, $n_\perp$) [**Fig. 5(b)**]. In literatures, the RI is ~1.9 (h-BN, in-plane), ~1.5 (h-BN, out-of-plane), and ~2.2 (c-BN), in the visible-to-near-infrared wavelength range.[40, 41] To mention, for *a*-BN the RI is much lower (~1.37 at 633 nm).[38] Also, for h-BN, ideally the out-of-plane RI should be lower than the in-plane RI, due to its 2D nature.[40] However, we observed the opposite trend (i.e. $n_\parallel < n_\perp$)[40]. This could be useful for the precise design of UWBG based optoelectronic (especially photonic) devices.[42]

Moreover, interfacial thermal transport between the UWBG semiconductors play an important role in thermal management applications while fabricating electronic devices. Therefore, we measured the thermal boundary conductance ($G$) and the cross-plane thermal conductivity ($k_\perp$) by using the optical pump-probe method, frequency domain thermo-reflectance (FDTR). An electro-optic modulator induced a sinusoidal intensity modulation on the 488-nm pump continuous wave laser, creating a periodic heat flux on the sample surface.[43] An unmodulated, 532-nm continuous wave probe laser monitored the surface temperature through a change in surface reflectivity. We compared the calculated phase lag of the sample surface temperature, induced by a periodic heat source at the sample surface, against the measured phase lag of the balanced probe beam (measured with respect to the reference signal from the lock-in amplifier).[44]

The sample is modeled as a three-layer system, where each layer includes the volumetric heat capacity $c_p$, $k_\perp$ and in-plane thermal conductivity ($k_\parallel$), layer thickness, and the thermal boundary conductance $G_1$ and $G_2$ [inset of **Fig. 5(c)**]. Au was chosen as a transducer layer to maximize the coefficient of thermo-reflectance at the probe wavelength. The measurement of individual material physical properties was performed as an inverse problem, minimizing the error between the calculated phase and the measured lock-in phase data via a nonlinear least-squares algorithm.[41] A more comprehensive description of solving this equation is detailed by Schmidt et al.[42] An example of the phase vs. frequency data obtained from FDTR of an average of three runs acquired on one spot location is provided [**Fig. 5(c)**]. The data is in good approximation to the best-fit curve obtained from solving the heat diffusion equation. Furthermore, sensitivity analysis is also implemented to determine which parameters can be fit together. The cross-plane thermal conductivity ($k_\perp$) of BN is the parameter of interest and it is most sensitive at higher frequencies [**Fig. 5(d)**]. The thermal boundary conductance ($G_1$) between Au and BN, $G_2$ between BN and diamond, and the thickness of BN are also most sensitive at higher frequencies. Thus, the focus is primarily on fitting the $k_\perp$ of BN and $G_2$. From the average of three runs on three separate spot locations on the sample (i.e., a total of nine measurements), we determined a $k_\perp$ for BN of 1.53 ($\pm$ 0.77) W/mK, and thermal conductance of the BN/SCD interface of $G_2$ = 20 ($\pm$ 2) MW/m$^2$K. These values are slightly lower than the bulk single-crystalline h-BN (4.8 W/mK and 60 MW/m$^2$K).[45] Brown et al., obtained thermal boundary conductance of 33.7 MW/m$^2$K for h-BN grown on Cu foil by CVD.[46] Glavin et al., obtained a much lower thermal conductance of a-BN (between 1-5 MW/m$^2$K) grown by PLD, by using different metal contacts.[47] Considering the next-generation high power electronics, information on the thermal conductivity and interfacial thermal boundary conductance of BN/diamond is of paramount importance due to its critical role in heat dissipation, thermal management, high-temperature operation, power density handling, and overall device reliability. These factors collectively contribute to the efficient and reliable operation of UWBG devices across a wide range of applications.

### III. CONCLUSIONS

To summarize, BN thin films were successfully grown on (100) diamond single crystals using PLD. Detailed structural analyses have confirmed the growth of these thin films. We investigated several properties through measurements of optical refractive index, magnetization, and thermal conductivity. While these properties are appealing, further endeavors are required to grow BN/diamond heterostructures with high-quality electronic interfaces in the future. Nonetheless, given the pressing demand for in-situ integrations with clean interfaces in high-power UWBG semiconductor electronics, our demonstration holds potential for fabricating BN/diamond devices suitable for diverse applications.

## ACKNOWLEDGMENTS

This work was sponsored partly by the Army Research Office and was accomplished under Cooperative Agreement Number W911NF-19-2-0269. The views and conclusions contained in this document are those of the authors and should not be interpreted as representing the official policies, either expressed or implied, of the Army Research Office or the U.S. Government. The U.S. Government is authorized to reproduce and distribute reprints for Government purposes notwithstanding any copyright notation herein. It was partly supported as part of ULTRA, an Energy Frontier Research Center funded by the U.S. Department of Energy, Office of Science, Basic Energy Sciences under Award No. DE-SC0021230, and in part by CHIMES, one of the seven centers in JUMP 2.0, a Semiconductor Research Corporation (SRC) program sponsored by DARPA. This work was also partly sponsored by the Department of the Navy, Office of Naval Research under ONR award number N00014-22-1-2357 and by the National Science Foundation Graduate Research Fellowship under Grant No. 1650114. This work was performed, in part, at the Cornell NanoScale Facility, a member of the National Nanotechnology Coordinated Infrastructure (NNCI), which is supported by the National Science Foundation (Grant No. NNCI-2025233). This work made use of the Cornell Center for Materials Research Shared Facilities, which are supported through the NSF MRSEC program (DMR-1719875). A. B. thanks Dr. Jianhua Li for kind help.

## AUTHOR DECLARATIONS

### Conflict of Interest

The authors have no conflicts to disclose.

### Author Contributions

A. B., R. V., and P. M. A. conceptualized the study. A. B., C. L., S. A. I., X. Z., H. K., T. G. and J. L. grew and characterized the films. A. B. P. and T. P. performed the FIB and electron microcopy. T. L. and Y. Z. carried out the optical measurement. G. A., J. C., E. J., and Z. T. measured thermal conductivity. A. G. B., M. R. N., E. G., B. B. P., and T. I. commented on the manuscript. All the authors discussed the results and contributed on the manuscript preparation.

## DATA AVAILABILITY

The data of this study are available from the corresponding author upon reasonable request.